\documentclass[aps,prl,floatfix,showpacs,twocolumn]{revtex4}
\usepackage{amsmath,amssymb}
\usepackage{graphicx}
\usepackage{epsfig}
\usepackage{psfrag}

\newcommand{\bm}{\boldsymbol}

\begin{document}

\hsize\textwidth\columnwidth\hsize\csname@twocolumnfalse\endcsname

\title{Spin and Valley Noise in Two-Dimensional Dirac Materials}


\author{Wang-Kong Tse, A. Saxena, D. L. Smith, and N. A. Sinitsyn}

\affiliation{Theoretical Division, Los Alamos National Laboratory, Los Alamos, New Mexico 87545, USA}

\begin{abstract}
We develop a theory for optical Faraday rotation noise in 
two-dimensional  Dirac materials.  
In contrast to spin noise in conventional semiconductors, we find that
the Faraday rotation fluctuations are influenced not only by spins but also the valley degrees of freedom
attributed to intervalley scattering processes. 
We illustrate our theory with two-dimensional transition metal
dichalcogenides and discuss  signatures of spin and valley noise in the Faraday noise power spectrum. We propose 
optical Faraday noise spectroscopy as a 
technique for probing both spin and valley relaxation dynamics in
two-dimensional Dirac materials. 
\end{abstract}

\maketitle

\textit{Introduction.--} 
The discovery of
two-dimensional (2D) Dirac materials has been a rapidly unfolding
trend recently \cite{balatsky-review}. 
Materials such as h-BN \cite{hBNPapers} and transition metal dichalcogenides, in particular MoS$_2$ semiconductor  \cite{MoS2Papers},  hold great promise for electronic and optical
applications \cite{NatNano}. 
An important feature of these \textit{Dirac semiconductors} 
is their intrinsic valley degrees of freedom that can couple to magnetic perturbations. 
2D Dirac materials are also promising
candidates for implementing spintronic devices due to their potential
in achieving long electron spin life-time and achieve non-dissipative
control of spin and valley currents via the quantum spin/valley Hall
effect \cite{dirac-she}. However, the spin dynamics of
near-equilibrium Dirac electrons in 2D MoS$_2$ is 
difficult to explore with a standard optical pump-probe setup because of strong excitonic effects \cite{nature-hanle}.   

The recently developed optical spin noise spectroscopy (SNS)
\cite{SN_Review} is an alternative 
technique to explore spin dynamics in semiconductors. It has been
used successfully to determine the spin coherence and spin
relaxation times in bulk GaAs  \cite{SN_Bulk}, quantum wells \cite{SN_QW,kavokin-13},
and quantum dots \cite{SN_QD}. Spin noise is studied via illumination of a 
linearly polarized light on a mesoscopic region of the sample.
The  Faraday rotation angle of the polarization axis of the
measurement beam is  proportional to the instantaneous total spin
polarization and can be detected  with a sensitivity reaching a single spin level \cite{single-spin} and picosecond time resolution \cite{high-band}. The 
noise power spectrum typically exhibits a peaked profile in a magnetic field transverse to the measurement axis, 
with the peak location yielding the g-factor and the broadening yielding the spin relaxation time. Importantly, SNS is a minimally invasive approach that does not require to artificially create a strongly nonequilibrium spin polarization.

In this Letter, we develop a theory of the optical spin noise 
spectroscopy for two-dimensional 
Dirac semiconductors. We take 2D MoS$_2$ as the prototypical example and study its spin noise dynamics within a Langevin equation
framework. We identify and discuss regimes of interest where fluctuations due to spin flip and inter-valley
scattering events predict measurable signatures in the Faraday
noise power spectrum from which information about spin and valley
relaxation processes can be inferred. Our main finding is that the Faraday rotation noise of Dirac electrons is sensitive to fluctuations in both spin and valley degrees of freedom. 
 Since the noise contribution from spins is sensitive to an external magnetic field, it should be possible to clearly separate spin and valley dynamics.

\textit{Faraday Rotation Fluctuations.--} We first establish a general  
relationship between the Faraday rotation fluctuations and the spin
noise in a 2D system of degenerate itinerant electrons.
To this end, we define the positive and negative helicity components of the
optical conductivity tensor as $\sigma_{\pm} = \sigma_{xx}\pm 
i\sigma_{xy}$, where $\sigma_{xx}$ and $\sigma_{xy}$ are the 
longitudinal and Hall conductivities of the 2D system. 
The real part $\mathrm{Re}\sigma_{\pm} =  \mathrm{Re}\sigma_{xx} \mp
\mathrm{Im}\sigma_{xy}$
corresponds to dissipative on-shell
electronic transitions whereas the imaginary part 
$\mathrm{Im}\sigma_{\pm} = \mathrm{Im}\sigma_{xx} \pm \mathrm{Re}\sigma_{xy}$ 
corresponds
to dissipationless virtual transitions. Throughout this Letter we adopt  `natural units' by normalizing optical conductivities 
by the speed of light $c$ so that  
$\sigma_{xx}, \sigma_{xy}$ are in units of the fine structure constant 
$\alpha \approx 1/137$. 
Analysis of the 
electromagnetic transmission problem through the 2D layer
\cite{Tse_TI} yields the Faraday rotation $\theta_{\mathrm{F}} = (\theta_{+}-\theta_{-})/2$, where
%
%
$\theta_{\pm} = -\tan^{-1}[2\pi \mathrm{Im}\sigma_{\mp}/(1+2\pi 
    \mathrm{Re}\sigma_{\mp})]$
is the phase of the positive and negative helicity component, 
respectively, of the electric field transmitted through the layer. It
is clear 
that $\mathrm{Im}\sigma_{\pm}$ 
contributes to 
circular birefringence and $\mathrm{Re}\sigma_{\pm}$
, in subleading
orders, to circular dichroism.  

 
The power of the optical spin noise spectroscopy lies in its
non-perturbative nature that allows it to probe the system under study 
at thermal equilibrium. This is achieved by tuning the light frequency
to a value that is smaller than the electronic band gap $E_{\mathrm{g}}$, with the
detuning $\omega_{\mathrm{d}} \ll E_{\mathrm{g}}$ so as to maximize the observed Faraday
rotation signal. Dissipative transitions are therefore forbidden with
$\mathrm{Re}\sigma_{\pm} \simeq 0$
, and the optical response is largely reactive and dissipationless.  
%
%
%

2D materials obeying the massive Dirac energy dispersion are described by
the Hamiltonian $H_0 = \hbar v \left(\tau
  k_x\hat{\sigma}_x+k_y\hat{\sigma}_y\right)+({\Delta}/{2})\hat{\sigma}_z$, where $v$ is the band velocity, $\Delta$ the band gap, 
$\bm{\hat{\sigma}}$ are the Pauli matrices describing the sublattice degrees of freedom, and $\tau = \pm 1$ denotes
the valley degrees of freedom $K, K'$. 
The valley degrees of freedom in addition to
spins are endowed with magnetic moments that can couple to
time-reversal breaking perturbations such as external magnetic
fields or circularly polarized light. Therefore the Faraday rotation
will be sensitive to the electron population differences between spins and valleys. 

To derive the Faraday rotation, we consider the situation with a nonequilibrium distribution of Fermi levels 
among the four possible quantum states $(K,\uparrow), 
(K,\downarrow), (K',\uparrow), (K',\downarrow)$ in the conduction
bands of the system, where $\uparrow,\downarrow$ denote the up and
down spins. 
Intra-valley spin-conserving scattering is fast compared to
intra-valley spin-flip or inter-valley scattering, so one can write
the corresponding Fermi levels for each spin and valley as 
$\epsilon_{F,s}^{\tau} = \epsilon_F+\delta \epsilon_{s}^{\tau}$, where 
$\tau = \pm 1$ corresponds to $K, K'$ and $s = \pm 1$ to
$\uparrow,\downarrow$,  $\epsilon_F$ is the equilibrium Fermi energy measured from the
conduction band edge, and $\delta \epsilon_{s}^{\tau}$ is the nonequilibrium Fermi
energy fluctuations for each spin $s$ and valley $\tau$. Such a nonequilibrium distribution of Fermi
energies is constrained by conservation of the total number 
of electrons, therefore $\delta \epsilon_{s}^{\tau}$ must satisfy an implicit condition given by 
$\sum_{s,\tau = \pm 1} \delta n_{s}^{\tau} = 0$ where 
$\delta n_{s}^{\tau}$ is the density fluctuations corresponding 
to $\delta \epsilon_{s}^{\tau}$. 

The optical conductivity due to such a nonequilibrium Fermi level
distribution is 
\begin{eqnarray}
\delta\sigma_{\pm} = \sum_{\tau = K,K'}\sum_{s = \uparrow,\downarrow}
\sigma_{\pm,s}^{\tau}(\epsilon_F+\delta \epsilon_{s}^{\tau}), 
\label{dsxy1}
\end{eqnarray}
where $\sigma_{\pm,s}^{\tau}$ refers to the conductivity for spin
$s$ and valley $\tau$ with the frequency label $\omega$ suppressed for clarity. We expand Eq.~(\ref{dsxy1}) up to first order in the small 
parameter $\delta \epsilon_{s}^{\tau}/\epsilon_F$ as appropriate for small
fluctuations. Noting that the two Hall contributions at the Fermi
level $\epsilon_F$ due to the $K,K'$ valleys cancel 
and using the expressions for $\theta_{\pm}$, 
%
we obtain 
the nonequilibrium Faraday rotation
\begin{eqnarray}
\theta_{\mathrm{F}} &=&
\frac{2\pi}{1+[2\pi\mathrm{Im}\sigma_{xx}(\epsilon_F)]^2} \nonumber \\
&&\sum_
{\tau = K,K'}\sum_{s =\uparrow,\downarrow}
\frac{\partial \mathrm{Re} 
    \sigma_{xy,s}^{\tau}(\epsilon)}{\partial
    \epsilon}\bigg\vert_{\epsilon = 
    \epsilon_F} \delta \epsilon_{s}^{\tau},  
\label{dFara}
\end{eqnarray}
%
where $\sigma_{xx}$ is the total longitudinal conductivity summed over 
all spins and valleys. 

\textit{2D transition metal dichalcogenides.--} Eq.~(\ref{dFara}) is 
applicable to massive Dirac fermion 2D systems or their multilayered 
derivatives.  
In particular, the large energy gap $E_{\mathrm{g}} \sim \mathrm{eV}$ of 2D 
transition metal dichalcogenides makes them especially suitable as prototypical 2D
Dirac materials amenable to study with SNS. Because of 
spin-orbit coupling, 2D transition metal dichalcogenides also exhibit
coupled spin and valley dynamics. The low-energy $k.p$ Hamiltonian near the
Brillouin zone corners $K,K'$ is given by  \cite{DiXiao_MoS2} $H = 
H_0+H_{\mathrm{SO}}$, where $H_0$ is the massive Dirac Hamiltonian
defined earlier 
and $H_{\mathrm{SO}} = -\lambda \tau
(\hat{\sigma}_z-1)\hat{s}_z/2$ is the spin-orbit coupling term with
${\hat{s}_z}$ the Pauli matrix describing the spin degrees of freedom
and strength $\lambda \sim 10\,-\,100\,\mathrm{meV}$. 




With the 
band gap given by $E_{\mathrm{g}} = \Delta-\lambda$, the frequency of
the probe laser beam is $\omega = \Delta-\lambda-\omega_{\mathrm{d}}$. 
The optical Hall and longitudinal conductivities are calculated using
the Kubo 
formalism and we find \cite{sup} 
%
\begin{eqnarray}
&&\mathrm{Re}\sigma_{xy,s}^{\tau} 
= -\frac{\tau \alpha}{4\pi}\frac{\Delta-s\tau\lambda}{\Delta-\lambda-\omega_{\mathrm{d}}} 
F_{s}^{\tau}(\epsilon_F,\omega_{\mathrm{d}}), \label{sxy}   \\
&&\mathrm{Im}\sigma_{xx,s}^{\tau} 
\label{sxx}
\\
&=&
-\frac{\alpha}{8\pi}\left\{\frac{2\epsilon_F+\Delta-s\tau\lambda}{ \Delta-\lambda-\omega_{\mathrm{d}}}\left[1+\left(\frac{\Delta-s\tau\lambda}{2\epsilon_F+\Delta-s\tau\lambda}\right)\right]^2
\right. \nonumber \\
&&\left.+\left[1+\left(\frac{\Delta-s\tau\lambda}{
        \Delta-\lambda-\omega_{\mathrm{d}}}\right)^2\right]F_s^{\tau}(\epsilon_F,\omega_{\mathrm{d}})
\right\},
\nonumber
\end{eqnarray}
for each spin and valley $s,\tau = \pm 1$ and 
$F_{s}^{\tau}(\epsilon_F,\omega_{\mathrm{d}}) = 
\mathrm{ln}\vert[2\epsilon_{F}-(s\tau-1)\lambda+\omega_{\mathrm{d}}]/[2\epsilon_{F}+2\Delta-(s\tau+1)\lambda-\omega_{\mathrm{d}}]\vert$. Equations
~(\ref{sxy})-(\ref{sxx})
are valid for our regime of interest at low temperatures $k_{\mathrm{B}}T \ll
\epsilon_F$ and for clean enough samples such that disorder broadening $\hbar/\tau \ll
\omega_{\mathrm{d}}$.  

%

To connect the Faraday
rotation fluctuations with experimental observables, we first express
Eq.~(\ref{dFara}) in terms of electron densities. The Fermi level
fluctuations can be related to the number density fluctuations as $\delta n_{s}^{\tau} = 
\nu_{s}^{\tau}(\epsilon_F)\delta\epsilon_{s}^{\tau}$, where $\nu_{s}^{\tau}(\epsilon)=
{(2 \epsilon+\Delta-\tau s\lambda)}/{4\pi (\hbar v)^2}$ is the density
of conduction band states per unit area. 
The nonequilibrium spin density 
fluctuations are given by $\delta s_z^K = (\delta n_{\uparrow}^K-\delta 
n_{\downarrow}^K)/2$ for valley $K$ and similarly for $K'$, whereas the 
valley density fluctuations are given by $\delta n^K = (\delta
n_{\uparrow}^K+\delta n_{\downarrow}^K)/2$, with $\delta n^{K'} = 
-\delta n^K$. Using these relations and substituting Eqs.~(\ref{sxy})-(\ref{sxx}) in
Eq.~(\ref{dFara}), we obtain the Faraday rotation fluctuations 
\begin{equation}
\theta_{\mathrm{F}} = \mathcal{L}_{s}S_z+\mathcal{L}_{v}N_{{v}}, 
%
\label{dFara2}
\end{equation}
%
where $S_z = (\delta s_z^K+\delta s_z^{K'})\mathcal{A}$ and $N_{v} = (\delta
n^K-\delta n^{K'})\mathcal{A}$ are respectively the total spin and
total valley polarization 
fluctuations over the cross-sectional area $\mathcal{A}$ of the
incident probe laser beam, 
and $\mathcal{L}_{s,v} =
(\mathcal{L}_{+}\pm\mathcal{L}_{-})/\mathcal{A}$ are the spin and
valley coupling coefficients respectively with 
$\mathcal{L}_{\pm} = \pm 8\pi\alpha(\hbar v)^2 (\Delta \mp
\lambda)/\{[\omega_{\mathrm{p}}^2-(2\epsilon_F+\Delta\mp
\lambda)^2](2\epsilon_F+\Delta\mp
\lambda)\{1+[2\pi\mathrm{Im}\sigma_{xx}(\epsilon_F)]^2\}\}$. 
Equation~(\ref{dFara2}) is a central result of this Letter. Importantly, valley 
fluctuations and spin fluctuations contribute on an equal footing to
the Faraday rotation in Dirac materials. 
This implies that the relaxation dynamics of both spins and valleys can be
optically probed using Faraday rotation spectroscopy. 

To determine the noise properties of the Faraday rotation, we now derive the kinetic equations governing the spin and valley
fluctuations. 
To this end, we first obtain an effective Hamiltonian
for conduction band states near the band edge. 
Using L\"{o}wdin's partitioning \cite{Lowdin} we find the following 
effective Hamiltonian for conduction band electrons with Fermi level 
$\epsilon_F \ll \Delta$ in an external field: 
%
\begin{equation}
H_{\mathrm{c}} =
\frac{\hbar^2k^2}{2m_e}+ \tau \frac{\hbar\Omega_{\mathrm{SO}}(k)}{2}  
\hat{s}_z +\tilde{g}(k) \mu_{\mathrm{B}}\bm{B}\cdot\bm{\hat{s}}, \label{effH}
\end{equation}
%
where  $m_e=(\Delta^2-\lambda^2)/(2\Delta v^2)$ is the
effective mass, $\tilde{g}(k) = g [1+(\hbar v
k)^2/(\Delta^2-\lambda^2)]$ is the renormalized g-factor and
$\Omega_{\mathrm{SO}}(k) =
2{\lambda}\hbar  (v k)^2/{(\Delta^2-\lambda^2)}$. Note that the spin-orbit coupling acts on conduction electrons as an effective out-of-plane magnetic field with opposite sign $\tau=\pm$ in different valleys. 


Using Eq.~(\ref{effH}) and the equation of motion for the spin-valley 
density matrix \cite{NonequilS}, we obtain the following kinetic equations for
the total spin fluctuations per valley $\bm{S}^{\tau}$ 
and total valley polarization fluctuations $N_{{v}}$ 
\begin{eqnarray}
&&\frac{\partial \bm{S}^{\tau}}{\partial t}+\tau \bm{S}^{\tau}\times 
\bm{\Omega}_{\mathrm{SO}}+\bm{S}^{\tau}\times \bm{\Omega}_{\mathrm{L}}  \nonumber \\
&=& -\frac{\bm{S}^{\tau}}{T_s}+\frac{\bm{S}^{-\tau}-\bm{S}^{\tau}}{T_v}
+\bm{\eta}_s^{\tau}+\bm{\eta}_v^{\tau},  \label{Seq} \\
&&\frac{\partial N_{{v}}}{\partial t} =
-\frac{N_{{v}}}{T_v}+\eta^{N}, \label{Neq} 
\end{eqnarray}
where $\bm{\Omega}_{\mathrm{SO}} =\Omega_{\mathrm{SO}}(k_F)\bm{z}$ is the effective out-of-plane  magnetic 
 field induced by spin-orbit coupling and $\bm{\Omega}_{\mathrm{L}} = 
 \tilde{g}(k_F)\mu_{\mathrm{B}}\bm{B}/\hbar$ at the Fermi level. The
 spin relaxation time $T_{s}$ \cite{Remark_T} captures the relaxation 
 of nonequilibrium spin fluctuations in one valley due to possible spin-flip 
scattering with magnetic defects \cite{MagI} as well as D'yakonov-Perel' and
Elliot-Yafet mechanisms \cite{SpinRelax}; whereas the valley
relaxation time $T_{v}$ describes the relaxation of valley
fluctuations due to inter-valley scattering from atomically-sized
non-magnetic impurities or electron-phonon interaction that induce momentum transfer on the order of inverse
lattice spacing \cite{Remark_ValleySpinFlip}. 
$\bm{\eta}_s^{\tau}$ and
$\bm{\eta}_v^{\tau}$ are the corresponding noise sources that describe spin fluctuations
$\bm{S}$ in valley $\tau$ due to spin and valley relaxation respectively, while $\eta^{N}$ describes fluctuations of the valley
polarization $N_v$. 
The different physical origin of these noise terms implies that
$\bm{\eta}_s^{\tau}$  is uncorrelated to 
$\bm{\eta}_v^{\tau}$ and $\eta^N$. The latter two can also be 
considered mutually uncorrelated if inter-valley scattering processes of 
up and down spins are statistically independent and equally probable. 
Due to the large (mesoscopic) number of electrons involved, the time 
correlators of these noise sources at thermodynamic equilibrium can be regarded as
$\delta$-function correlated, 
and are constrained by the Fluctuation-Dissipation Theorem 
to have the following form \cite{gardener,NonequilS}
\begin{eqnarray}
\langle (\eta_s^{\tau} (t))_{\alpha} (\eta_s^{\tau'}(t'))_{\beta}
\rangle &=& \delta_{\tau \tau'} \delta_{\alpha \beta}
\frac{2Dk_BT}{T_s} \delta(t-t'),  \label{corr1} \\
\langle (\eta_v^{\tau}(t))_{\alpha} (\eta_v^{\tau'}(t'))_{\beta}
\rangle &=& (\delta_{\tau \tau'} - \delta_{-\tau \tau'})
\delta_{\alpha \beta} \frac{2Dk_BT}{T_v} \delta(t-t'),  \nonumber \\ 
\label{corr2} \\
\langle \eta^{N}(t) \eta^{N}(t')  \rangle &=& \frac{4Dk_BT}{T_v} \delta(t-t'), 
\label{corr3} 
\end{eqnarray}
where $\alpha,\beta=x,y,z$, $T$ is temperature and $D$ 
is the density of conduction band states at the Fermi surface in the
observation area $\mathcal{A}$ per band and per
unit energy.

The Faraday rotation noise power follows from Eq.~(\ref{dFara2}): 
\begin{equation}
\langle \vert\tilde{\theta}_{\mathrm{F}}(\omega)\vert^2 \rangle=\mathcal{L}_{s}^2  \langle |\tilde{S}_z(\omega)|^2 \rangle+\mathcal{L}_{v}^2 \langle |\tilde{N}_v(\omega)|^2 \rangle,
\label{corr5}
\end{equation}
where $\tilde{S}_z(\omega)$ and $\tilde{N}_v(\omega)$ are the total 
spin and total valley polarization noise power with $
\tilde{X}(\omega) \equiv {\rm lim}_{T_m\rightarrow \infty} ({1}/{\sqrt{T_m}})\int_0^{T_m}\mathrm{d}t\,e^{i\omega t} X(t)$ where $T_m$ is the measurement time. 
Since $\bm{\eta}_s^{\tau}$, $\bm{\eta}_v^{\tau}$, $\eta^{N}$
are mutually uncorrelated, the dynamics of the valley polarization
$N_v$ decouples from that of the spins and can be readily obtained as 
\begin{equation}
{\langle |\tilde{N}_v(\omega)|^2 \rangle}=  \frac{4Dk_BT
  /T_v}{\omega^2 +1/T_v^2}.  
\label{charge}
\end{equation} 
%
\begin{figure}
\includegraphics[scale=.25]{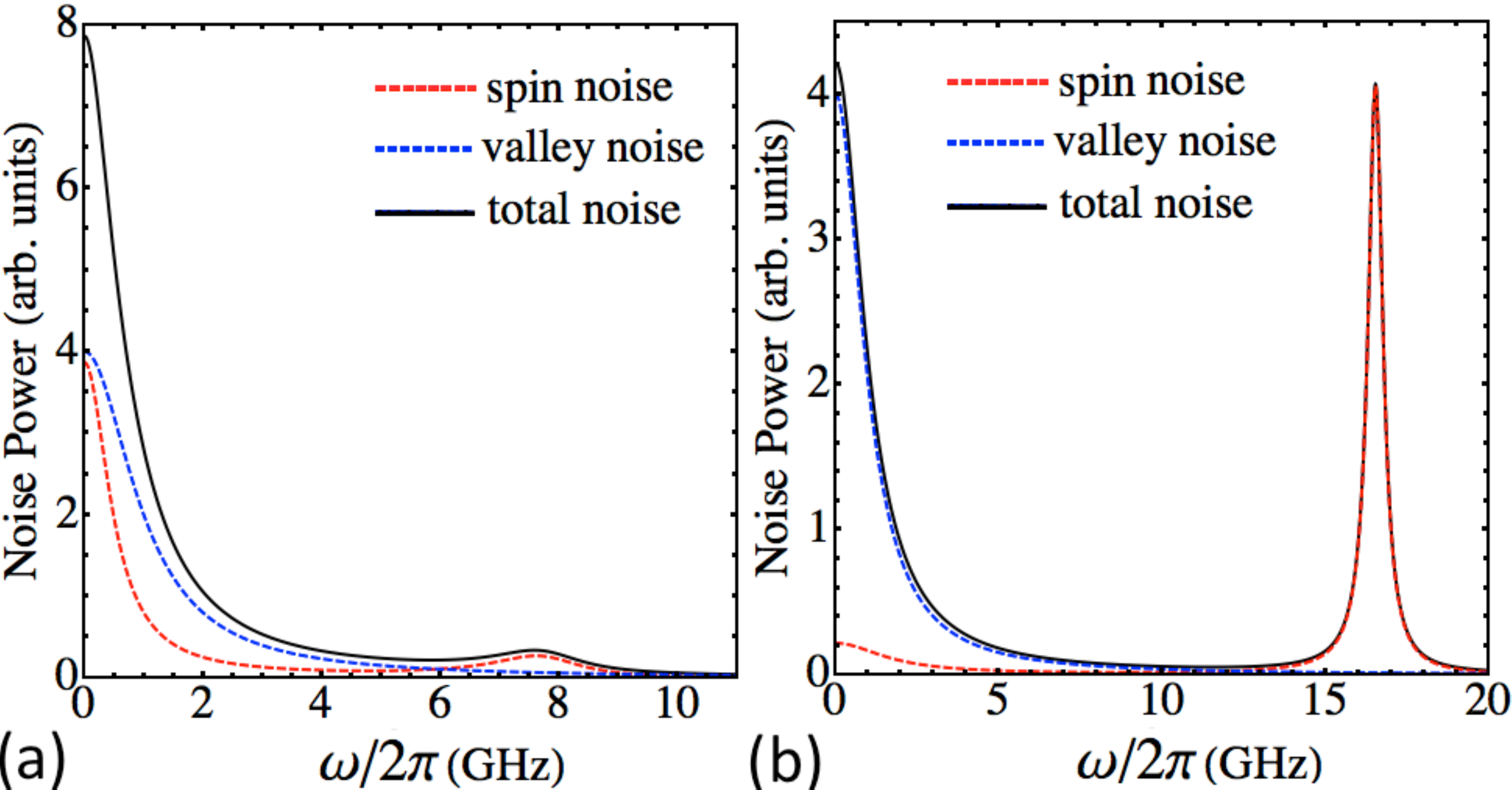}
\caption{\label{figure:P1n} (Color online) Total noise power spectrum (solid black)
  and its valley (dashed blue) and spin (dashed red) components at
  $\Omega_{\mathrm{eff}}> 1/T_s, 1/T_v$ with $T_s=31.42$ns, 
  $T_v=3.14$ns, and $\Omega_{\rm SO}/2\pi=7$GHz. (a) $\Omega_{\rm L}
  /2\pi =3.5$GHz ($\Omega_{\rm L}< \Omega_{\rm SO}$), and (b)
  $\Omega_{\rm L} /2\pi = 15$GHz ($\Omega_{\rm L} > \Omega_{\rm SO}$). Faraday rotation is assumed equally sensitive to valley and spin ($\mathcal{L}_{-}=0$). }
 \label{noise2}
\end{figure}
The valley dynamics is therefore described by a single Lorentzian noise power peak
centered at zero frequency that is insensitive to the 
applied magnetic field, and the valley relaxation time $T_v$ can be 
extracted from the width of the peak. For the spin dynamics, the
Langevin equation for spins Eq.~(\ref{Seq}) corresponds to an 
Ornstein-Uhlenbeck 
process \cite{gardener} with the spin noise correlators given by 
\begin{equation}
\langle \tilde{S}_{\alpha}^{\tau} (\omega) \tilde{S}_\beta^{\tau' } (-\omega) \rangle = \left( \frac{1}{{\bf A}-i\omega} {\bf G}   \frac{1}{{\bf A}^{\rm {\bf T}}+i\omega} \right)_{\tau \alpha, \tau' \beta},
\label{gard}
\end{equation}
where 
${\bm A}$ is the relaxation time matrix $A_{\tau \alpha, \tau'  \beta}
=- \delta_{\alpha \beta} \left( \delta_{\tau \tau'} /T_s
  +(\delta_{\tau \tau'}-\delta_{\tau,-\tau'})/T_v  \right)+
\delta_{\tau \tau'}\left( \tau \Omega_{\mathrm{SO}}
  \varepsilon_{\alpha z \beta} +\Omega_{\mathrm{L}}
  \varepsilon_{\alpha x \beta} \right)$ (here $\varepsilon_{ijk}$ is the Levi-Civita symbol) 
and  
$G_{\tau \alpha, \tau'  \beta}= \delta_{\alpha \beta} 2D k_B T \left( \delta_{\tau \tau'} (1/T_s+1/T_v) -\delta_{\tau -\tau'}/T_v \right)$.  
Equation~(\ref{gard}) can be evaluated analytically in closed form 
\cite{sup}. 


\begin{figure}
\includegraphics[scale=.23]{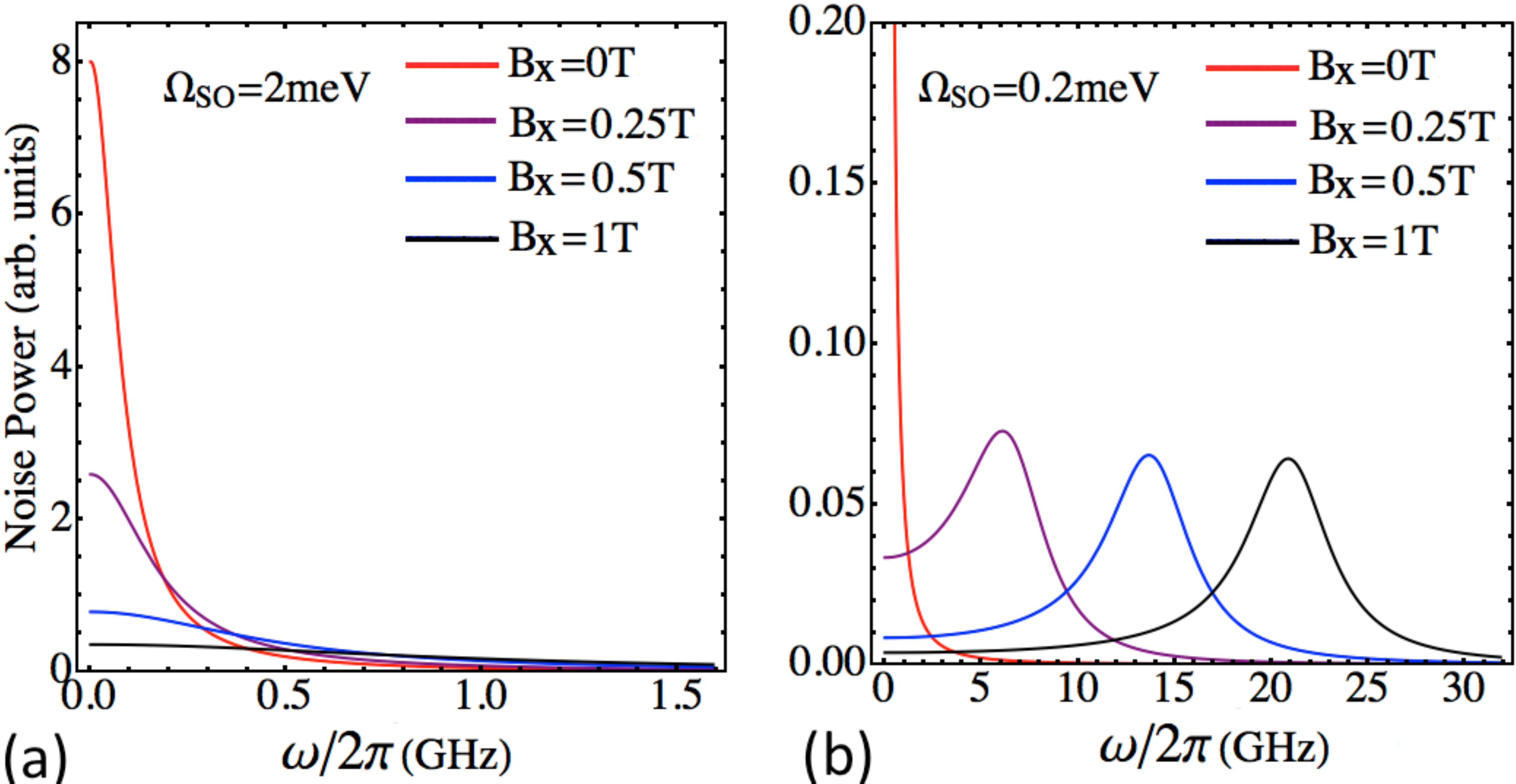}
\caption{\label{figure:P1n} Spin noise power spectrum of
  MoS$_2$ at different values of the external in-plane magnetic field
  (along x) for $T_v \ll T_s$. (a) $\hbar\Omega_{\rm SO}=2$meV, (b) $\hbar\Omega_{\rm SO}=0.2$meV. Values of parameters used are: g-factor $g=2$, $T_s=2$ns, $T_v=1$ps.}
 \label{noise3}
\end{figure}


The 2D transition metal dichalcogenide MoS$_2$ has a band gap in the 
optical spectrum $\Delta =1.7$eV and spin-orbit  
coupling $\lambda =75$meV \cite{DiXiao_MoS2}. 
For typical Fermi energies $\epsilon_F \sim 
10\,\mathrm{meV}$ near the band edge, detuning $\omega_{\mathrm{d}} \sim 10\,\mathrm{meV}$,
and laser spot size $\mathcal{A} \sim 1\,\mu\mathrm{m}^2$, we have 
total spin and valley polarization fluctuations $\sqrt{\langle |S_z|^2
\rangle}, \sqrt{\langle |N_v|^2 \rangle} \sim 200$ in the observation region with a total
number of electrons $\sim 10^4$. The Faraday rotation fluctuations is approximately equally 
coupled to the valley and spin degrees of freedom with $|\mathcal{L}_{-}/\mathcal{L}_{+}| 
\approx 0.1$ and the root-mean-squared Faraday angle fluctuations $\sqrt{\langle\vert\theta_{\mathrm{F}}\vert^2\rangle}  \sim
1\mu\mathrm{rad}$, which is within the state-of-the-art measurement
capabilities. 

In MoS$_2$, the spin-orbit splitting $\hbar\Omega_{\mathrm{SO}}\sim
\epsilon_F \lambda/\Delta \approx 0.5\,\mathrm{meV}$ 
corresponds to a strong magnetic field of $\sim
8\,\mathrm{T}$, which strongly favors out-of-plane spin alignments and
hence resists Larmor precession due to the applied in-plane field. 
Given that the spin relaxation time is 
expected to be long $T_s \sim 10$ns \cite{SpinRelax},  the behavior of the noise
spectrum will then critically depend on the valley scattering time.
Inter-valley scatterings are seen by electron spins as random sign
changes of 
the spin-orbit field ${\bm \Omega_{\mathrm{SO}}}$, so depending on the scattering rate this field may or may not be effectively averaged to zero.
Our studies of Eq.~(\ref{gard}) have identified three basic regimes \cite{sup}:




(i) {\it Slow inter-valley scattering:} $1/T_{v} \ll \Omega_{\rm SO}$. In this
regime, the Faraday rotation noise power spectrum consists of the valley noise 
component given by Eq.~(\ref{charge}) and spin noise components (Fig.~\ref{noise2}).  
At small fields $\Omega_{\mathrm{L}} \ll 
\Omega_{\mathrm{SO}}$, 
the spin noise part of the power spectrum is 
a Lorentzian peak centered at zero frequency 
 \begin{equation}
\langle |\tilde{S}_z(\omega)|^2 \rangle=  \frac{ 4Dk_BT/T_{s} }{\omega^2+1/T_{s}^2}. 
\label{spinnoise2}
\end{equation}
%
A finite frequency peak starts to emerge at
$\Omega_{\rm L}\sim \Omega_{\mathrm{SO}}$ [Fig.~\ref{noise2}(a)]. It is centered near the effective spin precession rate 
$\Omega_{\mathrm{eff}} =
\sqrt{\Omega_{\mathrm{L}}^2+\Omega_{\mathrm{SO}}^2}$. At  $\Omega_{\rm
  L}>\Omega_{\mathrm{SO}}$, the total noise power clearly consists of two peaks: one centered at zero frequency due to inter-valley scattering and the finite frequency peak
due to spin precession [Fig.~\ref{noise2}(b)]. In MoS$_2$,
$\Omega_{\mathrm{SO}}\sim\,$THz and the finite frequency peak cannot be resolved with currently accessible 100GHz bandwidth spectroscopy \cite{high-band}.

(ii) {\it  Moderately fast inter-valley scattering:} $1/T_{v} \lesssim
\Omega_{\rm SO}$.  
In this case, the finite frequency peak still appears only at large 
external fields, $\Omega_{\rm L}\sim\Omega_{\mathrm{SO}}$. 
However,  even a moderate in-plane magnetic field ($\lesssim 1\,{\rm T}$) induces a Dyakonov-Perel-type spin relaxation by making the directions
of the total field ${\bm  \Omega}_{\mathrm{SO}}+\bm{\Omega}_{\mathrm{L}}$ in $K$ and $K'$ valleys  non-collinear with each other. This results in a broadening of the zero-frequency peak 
as shown in Fig.~\ref{noise3}(a). 
The profile is generally non-Lorentzian, for which we have derived the relative amplitude of the correlator at zero and at finite magnetic 
fields as 
$\langle |\tilde{S}_z(\omega =0)|^2
\rangle_{\Omega_{\mathrm{L}}=0}/\langle |\tilde{S}_z(\omega = 0)|^2 \rangle_{\Omega_{\mathrm{L}}\ne
  0} = 1+2\Omega_{\mathrm{L}}^2T_s/(\Omega_{\mathrm{SO}}^2T_v)$, where $\langle |\tilde{S}_z(\omega =0)|^2
\rangle_{\Omega_{\mathrm{L}}=0}$. 
 By measuring the evolution of the peak maximum
with a changing external field, one can therefore obtain the combination of
parameters $\Omega_{\mathrm{L}}^2/(\Omega_{\mathrm{SO}}^2T_v)$ in
addition to the spin relaxation time $T_s$. 

(iii) {\it Fast inter-valley scattering:} $1/T_v >
\Omega_{\mathrm{SO}} $. In this case, the out-of-plane
spin-orbit field $\bm{\Omega}_{\mathrm{SO}}$ is quickly randomized. A moderate magnetic field is then  
sufficient to drive Larmor precession which displaces the spin noise peak from zero frequency to $\Omega_{\mathrm{L}} \sim \mathrm{GHz}$, as shown in
Fig.~\ref{noise3}(b). We find that the peak is then approximately Lorentzian near the
peak center $\Omega_{\mathrm{L}}$ 
with a spin relaxation time being renormalized by the fluctuating spin-orbit 
coupling $1/\bar{T}_s = 1/T_s+\Omega_{\mathrm{SO}}^2T_v/4$. This behavior can be observed with the 
state-of-the-art high bandwidth spin noise spectroscopy \cite{high-band}. 

{\it In conclusion,} we showed that the Faraday rotation noise of
Dirac electrons is sensitive to fluctuations of both the spin and
valley degrees of freedom. The noise power spectrum contains an
additional peak centered at zero frequency that is due to valley noise
and does not couple to external in-plane magnetic fields.
We also predict  that, due to spin-orbit splitting of electronic bands, a Larmor peak appears only in relatively strong external magnetic fields and 
its width depends on the spin-orbit splitting and inter-valley scattering rate. A moderate magnetic field $\sim 0.1-1\,$T is sufficient to strongly broaden the spin noise peak centered at zero frequency. If spin relaxation time is longer than $1\,$ns, this effect should be observable even without resorting to more complex high-bandwidth spectroscopy.


\begin{thebibliography}{30}
\bibitem{Remark_T} The small spin-orbit coupling corrections $\sim \lambda/\Delta$ to the 
  spin and valley relaxation times are neglected in our phenomenological relaxation time
  approximation. We expect that such corrections change only the
  quantitative but not the qualitative features we find.
\bibitem{Remark_ValleySpinFlip} Possible correction from inter-valley 
  spin-flip scattering, which requires the simultaneous breakdown of time-reversal symmetry and large momentum transfer, is negligible compared to
  intra-valley spin relaxation.
\end{thebibliography}
\end{document}